\documentclass[12pt]{article}
 \usepackage[dvips]{graphicx}
 \usepackage{amsmath}
 \usepackage{amsfonts}
 \usepackage{amssymb}
 \usepackage{longtable}

 \allowdisplaybreaks[4]     

 \renewcommand{\bf}{\bfseries}
 \renewcommand{\it}{\itshape}

 \newcommand{\showlabel}[1]{
   \label{#1}
 }
 \renewcommand{\d}{{\rm d}}     
 \newcommand{\td}[2]{\frac{{\rm d} {#1}}{{\rm d} {#2}}}
 \newcommand{\tdil}[2]{{\rm d} {#1} / {\rm d} {#2}}
 \newcommand{\p}{\partial}
 \newcommand{\pd}[2]{\frac{\partial {#1}}{\partial {#2}}}
 \newcommand{\pdil}[2]{\partial {#1} / \partial {#2}}
 \newcommand{\nn}{\nonumber}
 \newcommand{\er}[1]{(\ref{#1})}          

 \renewcommand{\L}{Lema\^{\i}tre}
 \newcommand{\LT}{\L-Tolman}
 \newcommand{\Rt}{\dot{R}}
 
 \newcommand{\Rh}{\hat{R}}
 \renewcommand{\th}{\hat{t}}
 \newcommand{\Mt}{\dot{M}}

\begin{document}

 \sffamily

 \title{The \L\ Model and the Generalisation of the Cosmic Mass}

 \author{
 Alnadhief H. A. Alfedeel%
    \thanks{ALFALN001@uct.ac.za}~
 and
 Charles Hellaby%
    \thanks{Charles.Hellaby@uct.ac.za}
 \\
 University of Cape Town,
 Rondebosch,
 7701,
 South Africa
 }

 \date{\today}

 \maketitle
 \begin{abstract}
   We consider the spherically symmetric metric with a comoving perfect fluid and non-zero pressure --- the \L\ metric --- and present it in the form of a calculational algorithm.  We use it to review the definition of mass, and to look at the apparent horizon relations on the observer's past null cone.  We show that the introduction of pressure makes it difficult to separate the mass from other physical parameters in an invariant way.  Under the usual mass definition, the apparent horizon relation, that relates the diameter distance to the cosmic mass, remains the same as in the \LT\ case.
 \end{abstract}

 \section{Introduction}

   The modern theory of cosmology is totally based on the general theory of relativity.  Underlying nearly all modern cosmological is the assumption of a homogeneous FLRW model.  Successful as it is, there are many unanswered questions about the nature of the matter content of the cosmos, as well as the relationship between observations of many discrete sources, which are necessarily averaged in some sense, and general relativity, which assumes the metric and the matter are described by smooth functions.  A significant complication arises from the fact that observers are looking down earth's past null cone, so we don't know the state of the universe at any one time or the history of any one worldline.

   When attempting to verify the question of homogeneity --- something we should do in the near future --- there are two distinct aspects.  Isotropy about us is relatively easy to check --- either the observations (at a given redshift) are (statistically) the same in different directions or they are not.  One doesn't need to know the spacetime evolution to check this.  But establishing radial homogeneity requires that we interpret observations in terms of a model that is not already assumed to be homogeneous.  The observations we make are affected by the bulk equation of state (which fixes the cosmic evolution), and also by any radial inhomogeneity that might be present.  These affect the shape of the past null cone and hence the time at which we observe each worldline.  The evolution of the sources properties must also be known, and theories of source evolution that have been validated or derived assuming a homogeneous model cannot be used to verify homogeneity.  

   The ``observational cosmology" series and related papers  \cite{ENMSW85,StElNe92a,StNeEl92b,StNeEl92c,SNME92,MaaMat94,MHMS96,AraSto99,AABFS01,ArRoSt01,RibSto03,AlIrRiSt07,HelAlf09,ASAB08} considered how the spacetime geometry could be determined from cosmological observations, starting with the classic by Kristian and Sachs \cite{KriSac66}, and using key concepts in \cite{Eth33,Tem38}.  An important construction is the use of observer coordinates, based on the past null cone.  Recent work on obtaining the metric of the cosmos from observations \cite{MuHeEl97,Hel01,Hel06,LuHel07,McCHel08} has focussed on the practicalities of a numerical procedure for converting observations into metric information.  To date this investigation has assumed spherical symmetry, as the key first step away from homogeneity towards the far more complex general case.  It was pointed out \cite{Hel06} that the apparent horizon --- the locus where the diameter distance is maximum --- has important observational consequences, that enable us to determine the cosmic mass at that radius.  This is due to a simple relationship between the maximum in the diameter distance, the cosmological constant, and the mass within a sphere of that size, that is unique to that distance.  This relationship allows us to check for systematic errors in the observational data, and partially correct them \cite{McCHel08}.  

   As early as 1933, \L\ \cite{Lem33} considered the general, diagonal, spherically symmetric metric, imposing a comoving, diagonal, matter tensor, but allowing the radial and tangential pressures to be different, and the cosmological constant to be non-zero.  He obtained the conservation equations, and from them obtained an expression for the mass.  He reduced the field equations to a system of first order differential equations (DEs) constraining the initial configuration and fixing the evolution.  He then went on to consider a large variety of very interesting cases and applications --- see \cite{Kra(Lem)97}.  This paper was well ahead of its time, and is little appreciated even today.%
    \footnote{Possibly the size of the paper was a hindrance to it's appreciation --- it contains enough material for several important papers.  Another aspect that possibly makes it difficult to penetrate at the start is the rapid introduction of a lot of new variables, the $\alpha$s and $\beta$s.  The initial diagonal metric seems to have signature $[++++]$, which means that the spatial metric components, $a_1^2$, $a_2^2$ \& $a_3^2$ are really negative (in the signature adopted later) so $a_1$ to $a_3$ are imaginary.  This makes the definition for the mass $m = - 4 \pi i \Phi$ rather peculiar, because $\Phi$ is imaginary, and it also introduces unexpected signs in $\Phi$'s definition (2.12).  Nevertheless, from section 3 onwards, the signature $[---+]$ and the notation are quite normal and the solution equations, e.g. (3.5), take an easily recognisable form.}

In 1964, Podurets \cite{MAP64} wrote out the field equations for a spherically symmetric ``star" consisting of a comoving perfect fluid (with isotropic pressure) and zero cosmological constant.  He solved them down to a slightly simpler set of evolution and constraint DEs, gave a brief thermodynamic interpretation of the equation relating pressure to the time derivative of mass, and presented only a sketch of how a numerical procedure would work.  In order to gain a basic interpretation of the equations, he showed how the Euler, continuity and Poisson equations are obtained in the Newtonian limit.

That same year, Misner and Sharp \cite{CWMDHS64} considered the same kind of star, pointing out it cannot include heat conduction or a flux of radiation.  They discussed the thermodynamics in some detail, and from this, they obtained the the same evolution and constraint DEs --- the relativistic Euler and continuity equations.  They defined a particle number and showed it is conserved.  They imposed an origin condition at zero radius, and a comoving boundary condition where the metric joins to a vacuum exterior.

Cahill and McVittie's \cite{CaMVittie70a} derivation of the spacetime is essentially the same as the above two, except they don't assume a diagonal matter tensor.  They provide several justifications for the choice of mass function, which we discuss below.  They go on to consider the perfect fluid, `negative mass shells',%
 \footnote{Since they assume $\rho$ \& $p$ are positive, these shells would appear to be the region beyond a maximum of $R$ in the spatial sections of constant $t$, where both $R$ and $M$ are decreasing towards a second origin.}
 and some specific metrics.

See \cite{Kra97} for an excellent review of this and many other inhomogeneous cosmological models.

 Given the significance of the apparent horizon result, it is important to check how general the relationship is. 
   The main content of this paper can be express as follows: In section two we will present the general \L\ metric, review how the solution of Einstein's field equation is reduced to a system of differential equations (DEs), and then show how one can construct  a \L\ model using appropriate numerical integration of the DEs, and finally we will show how the LTB and RW special cases are obtained. Section three will focus on how the definition of mass is affected by the introduction of non zero pressure and cosmological constant. In section four we will extend the apparent horizon relationship to the case of non-zero pressure, and discuss the past null cone with its observable quantities. We end this paper with the conclusions.

 \section{\L\ Metric}

   We consider an inhomogeneous, spherically symmetric spacetime, filled with a perfect fluid, in which the coordinates $x^{\mu} = (t,\,r,\,\theta,\,\phi)$ are comoving with the matter flow.  The metric may be written as
 \begin{align}\showlabel{metric}
   ds^2  = -e^{2\sigma} \, dt^2  + e^{\lambda} \, dr^2 + R^2 \, d\Omega^2 \;,
 \end{align}
 where $\sigma = \sigma(t,r)$, $\lambda = \lambda(t,r)$ are functions to be determined, $R = R(t,r)$ corresponds to the areal radius, and $d\Omega^2 = d\theta^2 + \sin^2 \theta \, d\phi^2$ is the metric of the unit 2-sphere.  The energy momentum tensor is given by:
 \begin{align}\showlabel{mT}
   T^{\mu\nu} = (\rho + p) \, u^{\mu} \, u^{\nu} + g^{\mu\nu}p \;,
 \end{align}
 where $\rho = \rho(t,r)$ is the mass-energy density of the perfect fluid, $p = p(t,r)$ is the matter pressure,  and $u^{\mu} = (e^{-\sigma}, 0, 0, 0)$ is the fluid four-velocity. This is less general than the full \L\ metric as we assume isotropic pressure.

\subsection{Field Equations} \showlabel{FEq}

   The EFEs, $G^{\mu\nu} = \kappa \,T^{\mu\nu} - g^{\mu\nu}\Lambda$ can be reduced to the following set of equations:
 \begin{align}
   e^{2\sigma} \, G^{tt} & = - \left( \frac{2 R''}{R } + \frac{R'^2}{R^2} - \frac{R'}{R} \, \lambda' \right)
      \, e^{-\lambda} + \left( \frac{\Rt^2}{R^2} + \frac{\Rt}{R} \, \dot{\lambda} \right)
      \, e^{-2\sigma} + \frac{1}{R^2} = \kappa \rho  + \Lambda ~.
      \showlabel{tt} \\ 
   e^{\lambda} \, G^{tr} & = \left( \frac{2 \Rt'}{R} - \frac{2 \Rt}{R} \, \sigma'
      - \frac{R'}{R} \, \dot{\lambda} \right) \, e^{-2\sigma} = 0 ~.
      \showlabel{zero} \\ 
   e^{\lambda} \, G^{rr} & = \left( \frac{R'^2}{R^2} + \frac{2 R'}{R} \, \sigma' \right) \, e^{-\lambda}
      - \left( \frac{2 \ddot{R}}{R} + \frac{\Rt^2}{R^2} - \frac{2 \Rt}{R} \, \dot{\sigma} \right)
      \, e^{-2\sigma} - \frac{1}{R^2} = \kappa p - \Lambda ~.
      \showlabel{rr} \\ 
   R^2 \, G^{\theta\theta} & = \left( \frac{R''}{R} + \frac{R'}{R} \, \sigma' + \sigma'' + \sigma'^2
      - \frac{R'}{2 R} \, \lambda'  - \frac{1}{2} \, \sigma' \, \lambda' \right) \, e^{-\lambda}
      \nonumber \\ 
   &~~~ + \left( \frac{\Rt}{R} \, \dot{\sigma} - \frac{\ddot{R}}{R} - \frac{1}{2} \, \ddot{\lambda}
      + \frac{1}{2} \dot{\lambda} \, \dot{\sigma} - \frac{\Rt}{2 R} \dot{\lambda}
      - \frac{1}{4} \, \dot{\lambda}^2 \right) \, e^{-2\sigma} = \kappa p - \Lambda ~.
      \showlabel{theta} 
 \end{align}
 Where the dot means a derivative with respect to $t$, and prime means a derivative with respect to $r$.  We use geometric units, $G = 1 = c$, so that $\kappa = 8 \pi$.  The conservation equations $\nabla_{\mu} T^{\mu\nu} = 0$ are:
 \begin{align}
   \frac{2 e^{2\sigma}}{(\rho + p)} \, \nabla_\mu T^{t\mu} & = \dot{\lambda}
      + \frac{2 \dot{\rho}}{(\rho + p)} + \frac{4 \Rt}{R} \; = 0
      \showlabel{grr} \\
   \frac{e^{\lambda}}{(\rho + p)} \, \nabla_\mu T^{r\mu} & = \sigma' + \frac{p'}{p + \rho} = 0 ~.
      \showlabel{gtt}
 \end{align}

   To solve these equations, we multiply Eq \er{tt} by $R^2R'$ and use Eq \er{zero} to eliminate the term that contains $\dot{\lambda}$, which produces 
 \begin{equation}\showlabel{ev}
   \frac{\partial}{\partial r} \left[ R + R \Rt^2 e^{-2\sigma} - R  R'^2 e^{-\lambda} - \frac{1}{3} \Lambda R^3 \right] = \kappa \rho R^2 R' \;.
 \end{equation}
 Multiplying Eq \er{rr} by $R^2\Rt$ and using Eq \er{zero} to eliminate the term that contains $\sigma'$, shows Eq \er{rr} can be rewritten as
 \begin{equation}\showlabel{Pr}
   \frac{\partial}{\partial t}\left[ R + R \Rt^2 e^{-2\sigma}-RR'^2 e^{-\lambda}-\frac{1}{3}\Lambda R^3\right] = -\kappa p R^2\dot{ R} \;.
 \end{equation}
 The term in square brackets is related to the total mass-energy of the system, $M$, interior to a comoving shell of constant $r$, and is normally defined by
 \begin{equation}
   \frac{2M}{R} = \Rt^2 e^{-2\sigma} - R'^2 e^{-\lambda} + 1 - \frac{1}{3} \Lambda R^2 \;.
   \showlabel{mR}
 \end{equation}
 The justifications for this will be discussed in section \er{M}.  With this definition, Eqs \er{ev} and \er{Pr} can be rewritten as:
 \begin{align}
   \kappa \rho & = \frac{2M'}{R^2 R'} ~,   \showlabel{density} \\
   \kappa  p & = -\frac{2\Mt}{R^2 \Rt} ~.   \showlabel{pressure}
 \end{align}
 These two relationships will be discussed in \S \ref{M}.  Eq \er{mR} may be rearranged as an evolution equation for the model
 \begin{align}
   \showlabel{Ev}
      \Rt = \pm e^{\sigma} \sqrt{\frac{2M}{R} + f + \frac{\Lambda R^2}{3}} ~, \\
 \intertext{where}
   \showlabel{f}
      f(t, r) = R'^2 e^{-\lambda} - 1~,
 \end{align}
 acts as the curvature term, or twice the total energy of the particles at $r$ (analogous to $f$ in the LT model).  However, the solution for $R(t,r)$ cannot be directly obtained from this, because of the unknown functions $\lambda$, $\sigma$ and $M$.  

   The metric variables $g_{tt}$ and $g_{rr}$ can be obtained by integrating Eqs \er{gtt} and \er{grr} 
as follows:
 \begin{align}
   \showlabel{si}
      \sigma & = \sigma_0(t) - \stackrel{\text{const}~t}{\int_{r_0}^r}  \frac{ p' \, \d r}{(\rho + p)}
      = \sigma_0 - \stackrel{\text{const}~t}{\int_{\rho_0}^\rho} \frac{(\pdil{p}{\rho})}{(\rho + p(\rho))} \, \d\rho~,
 \intertext{and}
   \showlabel{lamb}
      \lambda & = \lambda_0(r) - 2 \stackrel{\text{const}~r}{\int_{\rho_0}^\rho}
      \frac{\d\rho}{(\rho + p(\rho))}
      - 4 \ln \left( \frac{R}{R_0} \right)~,
 \end{align}
 where $\sigma_{0}(t)$ and $\lambda_{0}(r)$ are arbitrary functions of integration.  Typically, we would choose $r_0 = 0$ to be the origin, where $R(t, 0) = 0$.  Having solved \er{tt}-\er{rr} and \er{grr}-\er{gtt}, the $\theta$-$\theta$ field equation \er{theta} is now satisfied, since 
 \begin{align}
   R^2 \: G^{\theta\theta} & =
      \left( R \left( \frac{\sigma'}{4 R'} - \frac{\dot{\lambda}}{8 \Rt} \right)
      - \frac{1}{2} \right) e^{2\sigma} \: G^{tt}
      - \left( \frac{R e^{2\sigma}}{4 \Rt} \right) \p_t G^{tt} \nn \\
   &~~~~~~
      + \left( R \left( \frac{\sigma'}{4 R'} - \frac{\dot{\lambda}}{8 \Rt} \right)
      + \frac{1}{2} \right) e^{\lambda} \: G^{rr}
      + \left( \frac{R e^{\lambda}}{4 R'} \right) \p_r G^{rr} \nn \\
   &~~~~~~
      + \left( \frac{e^{2\sigma}}{4 \Rt} \left( \frac{R \lambda'}{2} - R \sigma'  - R'\right)
         + \frac{e^{\lambda}}{4 R'} \left( \frac{R \dot{\lambda}}{2} - R \dot{\sigma}  + \Rt \right) \right) G^{tr} \nn \\
   &~~~~~~
      + \left( \frac{R e^{\lambda}}{4 R'} \right) \: \p_t G^{tr}
      - \left( \frac{R e^{2\sigma}}{4 \Rt} \right) \: \p_r G^{tr}~,
 \end{align}
 and to get the right hand side of \er{theta} we must also add
 \begin{align}
   0 & =
      \frac{\kappa R e^{2 \sigma}}{4 \Rt} \: \nabla_{\nu} T^{t\nu} 
      - \frac{\kappa R e^{\lambda}}{4 R'} \: \nabla_{\nu} T^{r\nu}  ~.
 \end{align}

 \subsection{Constructing a \L\ Model}
 \showlabel{alg}

   To actually generate a \L\ model in the general case requires a numerical integration of the DEs.  In order to do that, we first need to specify the free functions, and then integrate the DEs along constant $t$ or $r$ paths.  Below is a summary of the procedure.
 \begin{itemize}
 \item   Choose an initial time slice $t_0$, and make a gauge choice $R(t_0, r) = R_0(r)$, for example $R_0 = r$.  This means $R_0'$ is also known.
 \item   Specify $\rho_0(r) = \rho(t_0, r)$ on that initial time slice.
 \item   Select a specific equation of state, $p = p(\rho)$, thus giving $p_0 = p(t_0,r)$, on the initial surface.
 \item   Integrate Eq \er{density} along constant $t = t_0$
 \begin{align}
   M_0 = \stackrel{\text{const}~t}{\int_{r_0}^r} \frac{\kappa \rho_0 R_0^2 R_0'}{2} \, \d r~,
 \end{align}
 to calculate $M_0 = M(t_0, r)$ everywhere on the initial time slice.  Alternatively, one may specify $M_0(r)$ and determine $\rho_0(r)$ from it.
 \item   Choose $\lambda(t_0, r)$, thereby fixing the free function $\lambda_0(r)$.  Note that by \er{f} this is equivalent to choosing the geometry $f_0(r)$ on the initial surface, so $\lambda_0(R) = 0$, $e^\lambda_0 = 1$ is not necessarily a good choice.  One may instead prefer to choose $f_0(r)$ and then calculate $\lambda_0(r)$ from $e^{\lambda_0} = R_0'^2/(1 + f_0)$.
 \item   Choose $\sigma(t, r_0)$ along the central worldline $r_0 = 0$, which we can take to be identically $\sigma_0(t)$.  This relates the time coordinate to the central observer's proper time.
 \item   Integrate \er{si} along $t = t_0$ to get $\sigma(t_0, r)$.
 \end{itemize}
 We have now determined all the functions on the initial time surface, which means we can obtain their $r$ derivatives too.  We next wish to evolve the metric forwards in time, along the worldlines of constant $r$.  This can be done as follows: 
 \begin{itemize}
 \item   Equation \er{Ev} gives $\Rt$ everywhere on our initial time slice.
 \item   The $\Mt$ DE can be obtained from Eq \er{pressure} as
 \begin{align}
   \Mt = \frac{- \kappa p \, \Rt R^2}{2} ~.
 \end{align}
 \item   Eliminating $\sigma'$ between Eqs \er{gtt} and \er{zero}, and then substituting for $\dot{\lambda}$ from \er{grr}, produces the following DE for $\dot{\rho}$
 \begin{align}
   \dot{\rho} = - p' \, \frac{\Rt}{R'}
      - (\rho + p) \left[ \frac{\Rt'}{R'} + \frac{2 \Rt}{R} \right] ~.
      \showlabel{rhodot}
 \end{align}
 \item   Having chosen the equation of state\footnote{More general equations of state would require a slightly different procedure} $p = p(\rho)$, an equation for $\dot{p}$ follows,
 \begin{equation}
   \dot{p} = \frac{dp}{d\rho} \, \dot{\rho}  ~.
 \end{equation}
 \item   Eq \er{grr} provides a DE for $\dot{\lambda}$, which may be used as is, or combined with \er{rhodot} to give
 \begin{align}
   \dot{\lambda} = \frac{2}{R'} \left( \frac{ p' \Rt}{(\rho + p)} + \Rt' \right)~.
 \end{align}
 \item   Having these 5 coupled DEs, and the initial values on $t = t_0$, we can create a numerical procedure to integrate the DEs in parallel, thus giving $R(t,r)$, $M(t,r)$, $\rho(t,r)$, $p(t,r)$ and $\lambda(t, r)$ everywhere.  Eq \er{density} can be used as a cross-check on the results.  Note that the spatial derivatives $\Rt'$, $p'$ and $M'$ will be needed at each step.
 \item   Finally, $\sigma(t,r)$ is obtained from Eq \er{si} by integrating along each slice of constant $t$.
 \end{itemize}
 Notice that we have chosen 4 functions, $R_0(r)$, $\rho_0(r)$, $\lambda_0(r)$ and $\sigma_0(t)$, as well as the equation of state $p(\rho)$.

 \subsection{Special Cases}

   The \L\ metric contains the \LT\ and FLRW metrics as special cases.  

   The spherically symmetric inhomogeneous dust cosmology of \LT\ is obtained when the pressure is zero.  Setting $p = 0$ in \er{gtt} gives $\sigma = \sigma_0(t)$, and we are free to set $\sigma = 1$ so that $t$ becomes the comoving proper time, i.e.\ the natural choice of time coordinate.  Eq \er{pressure} gives
 \begin{align}
   \Mt = 0 ~~~~\to~~~~ M = M(r) = M_0(r) ~.
 \end{align}
 Similarly, \er{grr} and \er{lamb} simplify to 
 \begin{align}
   \lambda & = \lambda_0(r) - 2 \ln \left( \frac{\rho}{\rho_0} \right) - 4 \ln \left( \frac{R}{R_0} \right) \\
      ~~~~\to~~~~ e^{\lambda/2} & = \frac{e^{\lambda_0/2} \rho_0 R_0^2}{\rho R^2}
      = \frac{e^{\lambda_0/2} R' M_0'}{M' R_0'}
      = \frac{R'}{(1 + f_0)}
 \end{align}
 where we used \er{density} and then \er{f}, so clearly $f = f_0(r) = f(r)$ too.  Therefore the evolution equation reduces to the familiar
 \begin{equation}
   \Rt^2 = \frac{2 M}{R} + f + \frac{\Lambda R^2}{3} ~.
 \end{equation}

   On the other hand, the standard cosmological model, the homogeneous RW metric, is obtained by setting to zero the spatial variation of any physical invariants.  Again $p' = 0$ allows us to choose $\sigma = 1$, and requiring the constant $t$ 3-spaces to have a constant curvature form requires $g_{ij} = S^2(t) \tilde{g}_{ij}(r)$, $i, j = 1,2,3$, with the canonical $r$ coordinate choice giving
 \begin{align}\showlabel{metricRW}
   ds^2 = -dt^2 + S(t) \left[ \frac{dr^2}{1 - k r^2} + r^2 (d\theta^2 + \sin^2\theta \, d\phi^2) \right] ~.
 \end{align}

 \section{Definition of Mass}
 \showlabel{M}

   In this section we will consider the justification for naming $M$ the gravitational mass felt at comoving radius $r$.  First we review the arguments given by some earlier authors, especially Cahill and McVittie, then we consider a geodesic deviation approach.

   \L\ \cite{Lem33} does not really justify calling $M$ the mass, other than an implied comparison with Newtonian equations.  Podurets \cite{MAP64} (with $\Lambda = 0$) merely says ``it is not difficult to see that" \er{mR} gives ``the total mass of matter contained in the interval $(0, r)$, the mass being defined in terms of the gravitational field created on the boundary", and then calls it a definition of mass.  Misner and Sharp \cite{CWMDHS64} (also using $\Lambda = 0$) write \er{density} in the form
 \begin{align}
   M & = \int_{V(r)} \rho \left( 1 + \Rt^2 e^{-2\sigma} - \frac{2 M}{R} \right)^{1/2} \d^3V ~,
 \end{align}
 pointing out that this $M$ contains contributions from the kinetic energy and the gravitational potential energy, as well as the matter density $\rho$.  (We note that the contributions $U^2 = \Rt^2 e^{-2\sigma}$ and $2M/R$ are twice the two energy values. We also note that the term $M$ appears in both sides, which causes a difficulty in interpreting this equation.)  

   Cahill \& McVittie \cite{CaMVittie70a}, assuming zero $\Lambda$, define the mass in terms of a Riemann component
 \begin{equation}\showlabel{MCM}
   M_{CM}(t,r) = \frac{R}{2} R^{\phi}{}_{\theta\phi\theta}
   = \frac{R}{2} \left( 1 + \frac{\Rt^2}{e^{2\sigma}} - \frac{R'^2}{e^{\lambda}} \right) ~.
 \end{equation}
 which involves only first metric derivatives, and is independent of any rescaling of the $t$ \& $r$ coordinates.  Their justifications are: 
 (i) if joined to a Schwarzschild exterior, $M$ must equal the exterior mass;
 (ii) the Bianchi identities lead to generalisations of \er{density} \& \er{pressure}, viz
 \begin{align}
   2 M' = \kappa R^2 (T^t_t \, R' - T^t_r \, \Rt) ~,~~~~~~
   2 \Mt = \kappa R^2 (T^r_r \, \Rt - T^r_t \, R') ~,
 \end{align}
 but the roles of the various terms in the interpretation are not discussed;
 (iii) the proper acceleration of $R$ on a comoving worldline is
 \begin{align}
   u^\mu \nabla_\mu (u^\nu \nabla_\nu R) = \frac{\kappa}{2} R T^r_r - \frac{M}{R^2}
   + R' e^{-\lambda} \lambda'
 \end{align}
 which contains a Newtonian-like gravitational force term $-M/R^2$, as well as the pressure acting on the comoving shell, and a term that's hard to interpret;
 (iv) they show that the mass flow vector
 \begin{align}
   J^\alpha = \frac{\sin\theta}{4 \pi \sqrt{|g_{\mu\nu}|}\;} (M', -\Mt, 0, 0) ~,~~~~~~ J^\alpha{}_{;\alpha} = 0 ~,
 \end{align}
 is conserved.%
 \footnote{Their (3.5) and (4.3) also seem to use a comma for a covariant derivative.}
 (For the case of the Reissner-Nordstrom metric, with mass and charge parameters $m$ and $q$, they find that 
 \begin{align}
   M = m - \frac{2 \pi q^2}{R} = m - 2 \pi \int_R^\infty E^2 \, R^2 \, \d R ~,
 \end{align}
 where $E$ is the electric field.  The two masses agree at $R = \infty$, but the effective gravitational mass $M$ decreases as the charge is approached, because the energy density in the $E$ field that is outside radius $R$ is not included in $M$.  As is well known, this energy density diverges close to the charge, so $M$ can go negative.)

   Aspects of the above arguments are clear and easy to follow, and for the $\Lambda = 0 = p$ case they are convincing.  It is not quite so easy to be certain we've properly understood the various relativistic corrections when $p \neq 0$.  A strong argument is that \er{density} is the same as in the LT model, and it applies at all times.  In other words, on every constant time 3-space, the gravitational mass is the integral of the density with a curvature (or gravity) correction, $\d^3V$ being the proper 3-volume element:
 \begin{align}
   M & = \int_{V(r)} \rho \,\sqrt{1 + f}\; \, \d^3V ~.   \showlabel{Mint}
 \end{align}
A similar treatment of \er{pressure} might seem to involve more complicated contributions from the gravity (curvature).  If we consider all the matter within the comoving sphere of constant $r$, then the work done by the pressure on the matter outside the boundary as it expands is $\d W = p \, A \, \d L$, where $A$ is the area of the boundary and $\d L$ its displacement.  Now if, following traditional Newtonian thinking, $\d L$ is taken to be the physical distance the boundary has moved, then this is
 \begin{align}
   \d W & = p \, A \, \td{}{t} \left( \int_0^r e^{\lambda/2} \, \d r \right) \, \d t \\
   & = \left( - \frac{2 \Mt}{\kappa R^2 \Rt} \right) \big( 4 \pi R^2 \big) \td{}{t} \left( \int_0^r \frac{R'}{\sqrt{1 + f}\;} \, \d r \right) \, \d t \\
   & = - \frac{\Mt}{\Rt} \td{}{t} \left( \int_0^r \frac{R'}{\sqrt{1 + f}\;} \, \d r \right) \, \d t ~.
 \end{align}
 which differs from the loss of mass-energy in the interior.  If instead we consider how much the boundary has expanded, not obviously the correct thing to do, since it has more to do with increase in surface area than radial distance moved, it gives the familiar Newtonian result:%
 \footnote{This thinking does not extend to \er{Mint}, which is unavoidably an integral over a volume, whereas \er{dWMtdt} is really about the boundary only.}
 $\d L = \Rt \, \d t$ leads to
 \begin{align}
   \d W & = - \Mt \, \d t ~.   \showlabel{dWMtdt}
 \end{align}

   Another problem arises once the cosmological constant is introduced, because Cahill \& McVittie's definition \er{MCM} then involves more than just the mass,
 \begin{equation}\showlabel{MCML}
   \frac{R}{2} \, R^{\phi}{}_{\theta\phi\theta} = M + \frac{\Lambda R^3}{6} ~.
 \end{equation}
 Consequently, in the following we investigate whether new invariant expressions can provide definitions of $M$, $\Lambda$ and other quantities independently.  
 
 \subsection{Geodesic Deviation Equation}

   We here look at another method that might clarify whether $M$ is the total mass-energy, based on the geodesic deviation equation (GDE).  The GDE measures the relative acceleration between two nearby geodesics, which is itself a measure of the tidal effects of gravity.  For a congruence of geodesics,
 \begin{equation}\showlabel{GD}
   \frac{\delta^2 \xi^{\mu}}{\delta \tau^2} = - R^\mu{}_{\nu\rho\lambda} \, U^{\nu} \, \xi^{\rho} \, U^{\lambda} ~,
 \end{equation}
 where $R^\mu{}_{\nu\rho\lambda}$ is the Riemann tensor, $U^{\nu}$ is the geodesic tangent vector $\tdil{x^\nu}{\tau}$, $\tau$ is proper distance or time along the geodesics, and $\xi^{\lambda}$ is the geodesic deviation vector.  Contracting this equation with $\xi_{\mu}$ produces the scalar
 \begin{equation}\showlabel{GDA}
   \mathcal{A}(\xi^\mu, U^\nu) = \xi_{\mu} \, \frac{\delta^2 \xi^{\mu}}{\delta \tau^2} = - R_{\mu\nu\rho\lambda} \, \xi^{\mu} \, U^{\nu} \, \xi^{\rho} \, U^{\lambda} ~,
 \end{equation}
 If $\xi^\lambda$ has unit magnitude, then $\cal A$ is the size of the relative acceleration of the chosen geodesics, and hence it is a measure of the strength of the tidal effects.  Now the Riemann tensor determines the tidal effects due to local as well as distant matter.  If one wishes to remove the effect of local matter, then one may consider the quantities
 \begin{equation}\showlabel{WD}
   - C^\mu{}_{\nu\rho\lambda} \, U^{\nu} \, \xi^{\rho} \, U^{\lambda} ~,
 \end{equation}
 and 
 \begin{equation}\showlabel{GDB}
   \mathcal{B}(\xi^\mu, U^\nu) = - C_{\mu\nu\rho\lambda} \, \xi^{\mu} \, U^{\nu} \, \xi^{\rho} \, U^{\lambda} ~,
 \end{equation}
 where $C^\mu{}_{\nu\rho\lambda}$ is the Weyl tensor.  Since we only wish to ``feel" the tides in a small neighbourhood, and we don't need to propagate the geodesics, we are free to choose $U^\nu$ and $\xi^\nu$ at will.%
 \footnote{More general measures of the curvature may be defined via the parallel transport equation, i.e. using the quantities $W_a \, R^a{}_{bcd} \, X^a \, Y^b \, Z^c$, but for the spacetimes under consideration, all the Riemann components with more than 2 different indices are zero.}

   In order to make these measures meaningful, we construct a set of canonical vectors using definitions that are invariant for any spherical metric.  In spherical symmetry, the ``areal radius" or ``curvature coordinate" is the metric component that multiplies the unit 2-sphere, $\d\Omega^2$.  In \er{metric} it is $R$.  Then the unit timelike vector that follows constant $R$ is,
 \begin{align}
   u^{\mu} u_{\mu} & = -1 ~,~~~~~~~~~~ u^{\mu} \nabla_{\mu} R = 0 ~~~~~\to~~~~~ \nn \\
   u^{\mu} & = \frac{e^{-\sigma-\lambda/2}}{\sqrt{e^{-\lambda} R'^2 - e^{-2\sigma} \Rt^2}\;} 
      \left( R', - \Rt, 0, 0 \right)~.
 \end{align}
 Using this, we define a canonical unit spacelike vector $v^{\mu}$ in the radial direction that is orthogonal to $u^{\mu}$ and satisfies
 \begin{align}
   v^\mu v_\mu & = 1 ~,~~~~~~~~~~ v^\mu u_\mu = 0 = v^\theta = v^\phi ~~~~~\to~~~~~ \nn \\
   v^{\mu} & = \frac{1}{\sqrt{e^{-\lambda} R'^2 - e^{-2\sigma} \Rt^2}\;} 
      \left( - \Rt e^{-2\sigma}, R' e^{-\lambda}, 0, 0 \right)~.
 \end{align}
 It is evident that the spacelike or timelike character of $u^\nu$ \& $v^\nu$ will flip if $(e^{-\lambda} R'^2 - e^{2\sigma} \Rt^2)\;$ changes sign.  We then define unit vectors in the $\theta$ and $\phi$ directions that are also orthogonal to  $u^{\nu}$ and $v^\nu$,  and tangent to the constant $R$ spheres,
 \begin{align}
   w^{\mu} w_{\mu} & = 1 ~,~~~~~~~~~~ u^{\mu} w_{\mu} = v^{\mu} w_{\mu} = 0 ~, \nn \\
   w^{\mu} \nabla_{\mu} R & = 0 ~~~~~\to~~~~~ w^{\mu} = \left( 0, 0, \frac{1}{R}, 0 \right)~,
 \end{align}
 and 
 \begin{align}
   z^{\mu} z_{\mu} & = 1 ~,~~~~~~~~~~ u^{\mu} z_{\mu} = v^{\mu} z_{\mu} = 0 ~, \nn \\
   z^{\mu} \nabla_{\mu} R & = 0 ~~~~~\to~~~~~ z^{\mu} = \left( 0, 0, 0, \frac{1}{R \sin\theta} \right)~.
 \end{align}
 Lastly, the radial incoming null vector $k^\mu$ is given by 
 \begin{align}
   k^{\mu} k_{\mu} = 0 = k^\theta = k^\phi ~~~~~\to~~~~~
   k^{\mu} = k \left( e^{\lambda/2}, - e^{\sigma}, 0, 0 \right) ~,
 \end{align}
 where $k$ is undetermined by the above conditions.  In addition, one could use the unit eigenvectors of the Einstein tensor $\left( G_{\mu\nu} - \ell g_{\mu\nu} \right) V^\mu = 0$ which are $e^{-\sigma} \delta^\mu_t$, $e^{-\lambda/2} \delta^\mu_r$, $R^{-1} \delta^\mu_\theta$, $(R \sin\theta)^{-1} \delta^\mu_\phi$, but it turns out these do not extend the range of independent scalars we can define.
 
   The scalars $\cal A$ \& $\cal B$, when combined with any canonical vectors, are determined solely by the metric, and therefore represent a property of the spacetime.  Below we use the notation
 \begin{align}
   AH = e^{-\lambda} R'^2 - e^{2\sigma} \Rt^2
      & = 1 - \frac{2 M}{R} - \frac{\Lambda R^2}{3}~, \\
   AHm = e^{-\lambda/2} R' - e^{\sigma} \Rt ~,~~~~~~
      & AHp = e^{-\lambda/2} R' + e^{\sigma} \Rt ~.
\end{align}
 Evaluating the scalars $\cal A$ and $\cal B$ using the various combinations of the above vectors gives:
 \begin{align}
   \showlabel{GE3}
      \mathcal{A}(u^\mu, v^\nu) & = \frac{2 M}{R^3} + \frac{\Lambda}{3}
      - \frac{\kappa}{2} (\rho + p) \\
   \mathcal{B}(u^\mu, v^\nu) & = \frac{2 M}{R^3} - \frac{\kappa \rho}{3} \\
   \showlabel{GE1}
      \mathcal{A}(u^\mu, w^\nu) = \mathcal{A}(u^\mu, z^\nu) & =
      \frac{\kappa \rho}{2} - \frac{\kappa(\rho + p) R'^2 e^{-\lambda}}{2 (AH)}
      - \frac{M}{R^3} + \frac{\Lambda}{3} \\
   \mathcal{B}(u^\mu, w^\nu) = \mathcal{B}(u^\mu, z^\nu) & = \frac{\kappa \rho}{6} - \frac{M}{R^3} \\
   \showlabel{GE6}
      \mathcal{A}(u^\mu, k^\nu) = \mathcal{A}(v^\mu, k^\nu) & =
      \frac{k^2 e^{\lambda+2\sigma} (AHm)}{(AHp)}
      \left[ \frac{2 M}{R^3} + \frac{\Lambda}{3} - \frac{\kappa}{2} (\rho + p) \right] \\
   \mathcal{B}(u^\mu, k^\nu) = \mathcal{B}(v^\mu, k^\nu) & = \frac{k^2 e^{\lambda+2\sigma} (AHm)}{(AHp)}
      \left[ \frac{2M}{R^3} - \frac{\kappa \rho}{3} \right] \\
   \showlabel{GE4}
      \mathcal{A}(v^\mu, w^\nu) = \mathcal{A}(v^\mu, z^\nu) & =
      \frac{\kappa p}{2} - \frac{\kappa (\rho + p) R'^2 e^{-\lambda}}{2 (AH)}
      + \frac{M}{R^3} - \frac{\Lambda}{3} \\
   \mathcal{B}(v^\mu, w^\nu) = \mathcal{B}(v^\mu, z^\nu) & =
      \frac{M}{R^3} - \frac{\kappa \rho}{6} \\
   \showlabel{GE2}
      \mathcal{A}(w^\mu, z^\nu) & = - \frac{2 M}{R^3} - \frac{\Lambda}{3} \\
   \mathcal{B}(w^\mu, z^\nu) & = \frac{\kappa \rho}{3} - \frac{2 M}{R^3} \\
   \showlabel{GE5}
      \mathcal{A}(w^\mu, k^\nu) = \mathcal{A}(z^\mu, k^\nu) & =
      - k^2 e^{\lambda+2\sigma} \frac{\kappa}{2} (\rho + p) \\
   \mathcal{B}(w^\mu, k^\nu) = \mathcal{B}(z^\mu, k^\nu) & = 0
 \end{align}
 We notice that all the $\cal B$s are multiples of $\kappa\rho/3 - 2M/R^3$, while the $\cal A$s are more diverse.  Combining the above equations, we can define the following 
 \begin{align}
   \mathcal{A}(v^\mu, w^\nu) - \mathcal{A}(u^\mu, w^\nu) & =
      \frac{2 M}{R^3} - \frac{2 \Lambda}{3} - \frac{\kappa (\rho - p)}{2} \\
   - \mathcal{A}(w^\mu, z^\nu) & = \frac{2 M}{R^3} + \frac{\Lambda}{3}
      \showlabel{ML} \\
   \mathcal{A}(v^\mu, w^\nu) - \mathcal{A}(u^\mu, w^\nu) - \mathcal{A}(u^\mu, v^\nu)
      & = \kappa p - \Lambda
      \showlabel{pL} \\
   - 2 \Big\{ \mathcal{A}(u^\mu, v^\nu) + \mathcal{A}(w^\mu, z^\nu) \Big\}
      & = \kappa (\rho + p)
      \showlabel{rhop} \\
   \frac{1}{3} \Big\{ \mathcal{A}(v^\mu, w^\nu) - \mathcal{A}(u^\mu, w^\nu)
      - \mathcal{A}(u^\mu, v^\nu) \Big\} - \mathcal{A}(w^\mu, z^\nu)
      & = \frac{2 M}{R^3} + \frac{\kappa p}{3}
      \showlabel{Mp} \\
   - \mathcal{A}(u^\mu, v^\nu) - 2 \mathcal{A}(w^\mu, z^\nu) - \mathcal{A}(v^\mu, w^\nu)
      + \mathcal{A}(u^\mu, w^\nu) & = \kappa \rho + \Lambda
      \showlabel{rhoL} \\
   \frac{1}{3} \Big\{ \mathcal{A}(v^\mu, w^\nu) - \mathcal{A}(u^\mu, w^\nu)
      + \mathcal{A}(u^\mu, v^\nu) - \mathcal{A}(w^\mu, z^\nu) \Big\}
      = \mathcal{B}(w^\mu, z^\nu) & = \frac{2 M}{R^3} - \frac{\kappa \rho}{3} ~.
      \showlabel{Mrho}
 \end{align}
 However, there are no combinations that give the value of just $M/R^3$, or $\kappa \rho$, or $\kappa p$ or $\Lambda$.  In the case of zero $\Lambda$, though, Eqs \er{ML}, \er{pL}, \er{rhoL} do give $M/R^3$, $\kappa p$ and $\kappa \rho$.  Similarly, in the case of zero pressure, Eqs \er{pL}, \er{rhop}, \er{Mp} do give $\Lambda$, $\kappa \rho$ and $M/R^3$.  This suggests that there is no definition of mass based on physical invariants in the general case.

   It appears from the results above, that we cannot separate the mass, the cosmological constant, the density and the pressure from each other, and so we cannot create a unique definition of mass based on geometric invariants of the metric in the general case. In contrast, if $p=0$ or $\Lambda=0$ (the LT  and Misner-Sharp-Podurets cases), then the remaining quantities are uniquely defined from these scalars.
This has important implications. The results of Cahill and McVittie do not generalise to the case of non-zero $\Lambda$, and the justification for calling the $M$ of \er{mR} the gravitational mass is weaker than was thought. It is based solely on comparing \er{density} and \er{pressure} with a few special cases. 

 \section{Past Null Cone, Apparent Horizon, and Cosmic Mass}

   We consider the past null cone (PNC) of the observation event $t = t_o$, $r = r_o = 0$.  For any quantity $Q(t,r)$, its value on our PNC will be indicated with a hat: $\hat{Q} = \hat{Q}(r) = Q(\hat{t}(r),r)$, or for expressions a square bracket with subscript ``$\wedge$" will be used.  For the metric \er{metric}, the path of an incoming radial light ray is given by
 \begin{align}
   \showlabel{tn}
   \td{\th}{r} & = -\frac{e^{\lambda/2}}{e^{\sigma}} ~,
 \end{align}
 and we define the solution that reaches $(t_o, r_o)$ to be $t = \th(r)$.  

   The redshift of comoving sources on that null cone is given by the ratio of the light oscillation periods $T$ measured at the observer, $o$, and the emitter, $e$,
 \begin{align}
   (1 + z)  = \frac{T_o}{T_e} ~.   \showlabel{zA}
 \end{align}
 For the two successive wavefronts passing through the events $B$ and $A$ or $D$ and $C$ respectively on a worldline of constant $r$, and a neighbouring one at $r + \d r$, the change in the light oscillation period $T$ over a distance $\d r$ is given by 
 \begin{align}\showlabel{Light}
   \d T & = \left. \td{\th}{r} \right|_C \, \d r - \left. \td{\th}{r} \right|_A \, \d r ~,
 \end{align}
 The first term on the right of Eq \er{Light} can be written as a Taylor expansion about $A$, so that
 \begin{align}
   \d T & = \left\{ \left. \td{\th}{r} \right|_A
      + \left. \pd{}{t} \left( \td{\th}{r} \right) \right|_A T +\left. \pd{}{r} \left( \td{\th}{r} \right) \right|_A (0)
      \right\} \, \d r
      - \left. \td{\th}{r} \right|_A \, \d r \\
   \frac{\d T}{T} & = \left. \pd{}{t} \left( \td{\th}{r} \right) \right|_A \,  \d r
 \end{align}
 Therefore Eq \er{Light} can be integrated down the PNC to give
 \begin{align}\showlabel{ToTe}
   -\ln \left( \frac{T_o}{T_e} \right) & = \int_0^{r_e} \pd{}{t} \left( \td{\th}{r} \right) \, \d r ~,
 \end{align}
 so that, for the \L\ metric, Eqs \er{tn} and \er{zA}, give the redshift as
 \begin{align}
   \ln(1 + z) = \int_0^{r_e} \pd{}{t} \left( \frac{e^{\lambda/2}}{e^{\sigma}} \right) \, \d r ~.
 \end{align}

   With any cosmological sources there are primary observable quantities such as redshift $z$, angular diameter $\delta$, apparent luminosity $\ell$, and number density in redshift space $n$.  Associated with each of $\delta$, $\ell$, and $n$ is a source property, true diameter $D$, absolute luminosity $L$, and mass per source $\mu$, which may evolve with time, and therefore vary with redshift.  These combine to give the luminosity distance $d_L$, the diameter distance $d_D$ and the mass density in redshift space, $\mu n$.  

   We define the diameter distance of a source as the ratio between the true diameter $D$ and the angular diameter $\delta$.  Consequently, in a spherical metric, it corresponds to the areal radius evaluated on the PNC, i.e.
 \begin{align}
   \frac{D}{\delta} = d_D = \hat{R} ~.   \showlabel{dD}
 \end{align}
 Given the absolute luminosity of a source $L$, and the apparent luminosity $\ell$ (or $m$ and $\tilde{m}$, the apparent and absolute magnitude), then the luminosity distance is
 \begin{align}
   d_L = \sqrt{\frac{L}{\ell}}\; \, d_{10} = 10^{(m - \tilde{m})/5} \, d_{10} ~.   \showlabel{dL}
 \end{align}
 where $d_{10}$ is 10 parsecs.  According the the reciprocity theorem \cite{Eth33,Penr66,Ell71}, the luminosity distance $d_L$ is related to the diameter distance $d_D$ by
 \begin{align}
   d_D = {d_L}{(1 + z)^2} ~.   \showlabel{rcprct}
 \end{align}

   Working in redshift space, $(z, \theta, \phi)$, let $n(z)$ be the density of sources, that is the number per steradian per unit redshift interval.  Suppose that there are $dN$ sources in solid angle $d\omega = \sin\theta \, d\theta \, d\phi$ between redshift $z$ and $z + dz$, and that $\mu(z)$ is the mean mass per source, then 
 \begin{align}
   d{\cal M} = \mu \, dN = \mu \, n \, d\omega \, dz ~.
 \end{align}
 The 3-d volume $d^3V$ of proper space that encloses these sources at the time of emission $t_e$, as measured by comoving observers $u^{\mu}$, is given by $d^3V = e^{\lambda/2} R^2 \, d\omega \, dr$, so that the mass in the fluid element is
 \begin{align}
   d{\cal M} & = \rho \, e^{\lambda/2} R^2 \, d\omega \, dr ~.
 \end{align}
 Therefore the relationship between $n$ and $\hat{\rho}$ is given by
 \begin{align}
   \mu \, n = \left[ \rho \, e^{\lambda/2} \, R^2 \, \td{r}{z} \right]_\wedge ~.
   \showlabel{munrho}
 \end{align}

   The apparent horizon is the locus where the observer's PNC reaches it's maximum areal radius (diameter distance), which we denote $z = z_m$, $\Rh = \Rh_m$, and obviously we have $[\tdil{\Rh}{r}]_m = 0$.  The total derivative of $\Rh$ along the ray is
 \begin{align}
   \td{\Rh}{r} = \left[ R' + \Rt \, \td{\th}{r} \, \right]_\wedge
 \end{align}
 and using Eqs \er{Ev} and \er{tn} to replace $\Rt$ \& $\tdil{\th}{r}$ shows
 \begin{align}
   \td{\Rh}{r} & = e^{\lambda/2} \left[ R' e^{-\lambda/2} - 
      \sqrt{\frac{2 M}{R} + R'^2 e^{-\lambda} - 1 + \frac{\Lambda R^2}{3}}\; \right]_\wedge = 0 \\
      & \longrightarrow~~~~~~ 6 M_m - 3 \Rh_m + \Lambda \Rh_m^3 = 0
      \showlabel{AH}
 \end{align}
 at $\Rh_m$.  This is exactly the same as the result in \cite{Hel06}, showing that the non-zero pressure does not affect the apparent horizon condition.  Note too that $M$ and $\Lambda$ appear in exactly the same inseparable combination as in \er{ML}, so \er{AH} can be written $\Rh_m^2 = -1/{\cal A}(w^\mu, z^\nu)|_m$.

 \section{Conclusions}
 \showlabel{Concl}

   We have considered the \L\ metric, which describes a spherically symmetric perfect fluid distribution in comoving coordinates, and its familiar reduction to a system of DEs.  We presented the solution in the form of an explicit algorithm for numerical calculation, clearly listing those functions that need to be specified on an initial surface, and on a chosen worldline.  
   
   The concept of the total gravitational mass within a given comoving sphere has been reviewed in the case of both pressure and cosmological constant being non-zero, and, in addition to the usual relativistic corrections whose role is hard to pin down, it was found that $\Lambda$ appears in the Cahill \& McVittie definition of $M$.  In seeking for alternative definitions or justifications, we considered a number of invariants of a spherical metric, based on the geodesic deviation equation and a set of canonical vector fields.  We found that the physical variables $M$, $\rho$, $p$ and $\Lambda$ always appear in combinations of two or more, and it is not possible to separate them using methods based on invariant properties of the metric.  Since $M$ is not just the integral of the density, but contains a contribution from the curvature, and since the relation between $\Mt$ and the work done by the pressure is not straightforward, we are therefore left with some ambiguity in the definition of the effective gravitational mass.

   We looked at the past null cone of a central observer, and how the areal radius (diameter distance) varies along it.  The apparent horizon relation found in \cite{Hel06}, that is the relationship between the maximum in the diameter distance $\Rh_m$, $\Lambda$, and the mass $M$ within that sphere --- the ``cosmic mass" ---  has been shown to hold good for the \L\ model.  This relationship, previously only shown for the zero pressure case, holds only at the apparent horizon, and is independent of any intervening inhomogeneity.  It provides a way of measuring the combination $(M_m + \Lambda \Rh_m^3/6)$ on gigaparsec scales.

   Now it is known \cite{MuHeEl97} that the \LT\ model can fit any given observational functions for diameter distance and number counts versus redshift.   A generalisation of this theorem to the \L\ model would seem to offer more flexibility, and thus the possibility of fitting other data.

 \section{Acknowledgements}
 \setcounter{secnumdepth}{2}

   CH thanks South Africa's National Research Foundation (NRF) for a grant.  AHAA thanks the NRF and the African Institute for Mathematical Sciences (AIMS) for bursaries.

 \appendix

 \section{Barotropic Equation of State}
 \showlabel{BEOS}

   For the case of the commonly used equation of state, $p = w\rho$, with $w$ the equation of the state parameter, the \L\ model has a more explicit solution.  Eq \er{si} and \er{lamb} can be integrated as follows
 \begin{align}
      \sigma - \sigma_0(t) & = - \stackrel{\text{const}~t}{\int_{r_0}^r}  \frac{ w \rho' \, \d r}{\rho(1 + w)}
      = \frac{-w}{(1 + w)} \ln \left( \frac{\rho}{\rho_{t0}} \right)
 \intertext{where $\rho_{t0} = \rho(t, r_0)$, and}
      \frac{\lambda}{2} - \frac{\lambda_0(r)}{2} & = - \stackrel{\text{const}~r}{\int_{t_0}^t}
      \frac{\rho' \, \d r}{\rho (1 + w)} - 2 \ln \left( \frac{R}{R_{0r}} \right)
      = \frac{-1}{(1 + w)} \ln \left( \frac{\rho \, R^2}{\rho_{0r} \, R_{0r}^2} \right)
 \intertext{where $R_{or} = R(t_0, r)$ etc, so that}
   \showlabel{esiglamw}
      e^{\sigma - \sigma_0} & = \left( \frac{\rho}{\rho_{t0}} \right)^{-w/(1 + w)} ~,~~~~~~
      e^{(\lambda - \lambda_0)/2} = \left( \frac{\rho \, R^2}{\rho_{0r} \, R_{0r}^2} \right)^{-1/(1 + w)} ~.
 \end{align}


 \end{document}